\documentclass[twocolumn,showpacs]{revtex4}
\usepackage{epstopdf}
\usepackage{graphicx}
\usepackage{dcolumn}
\usepackage{bm}

\def\gappeq{\mathrel{ \rlap{\raise.5ex\hbox{$>$}}
                      {\lower.5ex\hbox{$\sim$}}  } }
\def\lappeq{\mathrel{ \rlap{\raise.5ex\hbox{$<$}}
                      {\lower.5ex\hbox{$\sim$}}  } }

\begin{document}

\preprint{PRA}

\title{Transmission and Reflection of Bose-Einstein Condensates Incident on a
Gaussian Potential Barrier}

\author{A.M. Martin$^{1}$, R.G. Scott$^{2,3}$ and T.M. Fromhold$^{3}$}
\address{$^{1}$ School of Physics, University of Melbourne, Parkville,
Victoria 3010, Australia \\ $^{2}$ Department of Physics, University
of Otago, P.O. Box 56, Dunedin, New Zealand \\$^{3}$ School of
Physics and Astronomy, University of Nottingham, Nottingham NG7 2RD,
United Kingdom}
\date{\today}

\begin{abstract}
We investigate how Bose-Einstein condensates, whose initial state is
either irrotational or  contains a single vortex, scatter off a
one-dimensional Gaussian potential barrier. We find that for low
atom densities the vortex structure within the condensate is
maintained during scattering, whereas at medium and high densities,
multiple  additional vortices can be created by the scattering
process, resulting in complex dynamics and disruption of the atom
cloud. This disruption originates from two different mechanisms
associated respectively with  the initial rotation of the atom cloud
and the interference between the incident and reflected matter
waves.
%
We investigate how the reflection probability depends on the
vorticity of the initial state and on the incident velocity of the
Bose-Einstein condensate. To interpret our results, we derive a
general analytical expression for the reflection coefficient of a
rotating Bose-Einstein condensate that scatters off a
spatially-varying one-dimensional potential.
\end{abstract}

\maketitle

\section{INTRODUCTION}
Recently, several experiments have investigated the scattering of
Bose-Einstein condensates (BECs).
These experiments have included Bragg reflection in an optical
lattice \cite{Morsch}, reflection from optical \cite{mirror1} and
magnetic \cite{mirror2} mirrors, diffraction from a grating \cite{X}
and quantum reflection from a silicon surface \cite{Pasquini}. In
each case, interest has focused on the reflected component of the
BEC. For example, investigations of quantum reflection from a
silicon surface have revealed that inter-atomic interactions have a
dramatic effect on the internal structure of the atom cloud
\cite{Pasquini,preprint}. So far, reflection experiments have been
restricted to condensates whose initial state contains no dynamical
excitations. However, the methodology for creating and observing
vortices in BECs is well established
\cite{vortex1,vortex2,vortex3,vortex4} and, in the case of Bragg
reflection, numerical simulations predict that vortices and solitons
in the BEC's initial state strongly influence the subsequent
dynamics \cite{PRA1}. Previous theoretical work has also shown that
the presence of a vortex in the initial state can have a pronounced
 effect on the internal structure of a BEC that undergoes
classical reflection from a hard wall atom mirror
\cite{opposite1,opposite2}. To our knowledge, however, there has
been no consideration of the effect of vortices on BECs approaching
a potential barrier of {\it finite} width and height, where
quantum-mechanical tunneling is possible as well as reflection.
Quantum tunneling of BECs can be studied experimentally by using
sheets of laser light to create the potential barrier
\cite{Ketterle} and plays a crucial role in the dynamics and
macroscopic coherence of cold atoms in optical lattices \cite{MOT}.

In this paper, we investigate how  BECs scatter off a Gaussian
potential barrier, which allows  both reflection and
quantum-mechanical tunneling of the atoms. We use numerical
simulations of the Gross-Pitaevskii equation to make a detailed
study of how the strength of the inter-atomic interactions, the
energy of the incident BEC, and the vorticity of the initial state
affect the scattering properties of the condensate. Our simulations
reveal regimes in which dynamical excitations disrupt both the
reflected and transmitted atom clouds. In one regime, which we call
{\it rotational disruption}, the excitations originate from the
effect of the initial vortex on the scattering dynamics. By
contrast, the regime of {\it interferential disruption} \cite{Robin}
occurs both in the presence and absence of a vortex in the initial
state with the excitations being created from interference between
the incident and reflected matter waves.
We find that rotational disruption
arises when the time taken for the BEC to scatter is comparable
with, or exceeds, the rotational period of the vortex.
Interferential disruption occurs when the scattering time is greater
than the correlation time of the BEC, which is a measure of how
quickly the atom cloud responds to a perturbation. To interpret our
results, we derive a general expression for the reflection
probability of a BEC containing a vortex in terms of the reflection
probability of an irrotational BEC. We find that the vortex changes
the reflection probability by altering the distribution of incident
velocities at the barrier, due to the increase in the physical size
of the cloud and, more importantly, the circulation of the atoms.

%
%

To investigate the regimes of rotational and interferential
disruption we consider three BECs whose initial state contains a
single vortex: BEC $A$ with low atom density,
BEC $B$ with medium density and
BEC $C$ with high density.
We identify the effect of the initial vortex by comparing the
dynamics of these atom clouds with irrotational counterparts labeled
BECs $A_i$, $B_i$ and $C_i$.

The layout of the paper is as follows: in section II we specify the
parameters for each of the three BECs and describe the theoretical
model and computational techniques. In section III, we present our
numerical results, which show how low, medium, and high density BECs
scatter off the Gaussian potential barrier, both with and without a
vortex in the initial state.
In section IV, we derive a general expression for the reflection
probability of an irrotational BEC. We then use this expression to
derive an approximate analytical formula for the reflection
probability of a condensate whose initial state {\it does} contain a
vortex, impinging on the same scattering potential. We compare this
formula with reflection probabilities obtained directly from the
numerical simulations presented in section III and use it to provide
physical insight into the effect of vortices on the scattering
process. Finally, in section V, we summarize our results and propose
experiments to test them.
\section{SYSTEM PARAMETERS, THEORETICAL MODEL, AND COMPUTATIONAL TECHNIQUES}
Each BEC contains $N$ $^{23}$Na atoms of mass $m$ and is initially
confined by a harmonic trapping potential, $V_T(x,y,z)= m
[\omega_x^2 (x+\Delta x)^2 +\omega_y^2 y^2 + \omega_z z^2]/2$
centered at $(-\Delta x$,0,0). We consider trap frequencies
$\omega_x=\omega_y=50 \times 2 \pi $ rad s$^{-1} \ll \omega_z$ so
that the spatial width of the BEC
 is much smaller along the $z$-direction than
along the $x$- and $y$-directions. Consequently, the dynamics reduce
to two-dimensional motion in the $x-y$ plane. At time $t=0$, we
create an additional Gaussian potential barrier \cite{Ketterle}
\begin{eqnarray}
V_L(x)=V_0e^{-\frac{x^2}{\sigma^2}},
\end{eqnarray}
of width  $\sigma=1 \, \mu$m along the $x$-direction, by switching
on a far blue-detuned laser  beam that travels along the
$y$-direction and creates a sheet of laser light. The intensity of
the laser beam determines the barrier height, $V_0$, which we take
to be $6.2\,$peV, similar to recent experiments \cite{Ketterle}.
Simultaneously, we accelerate the BEC towards the Gaussian potential
by abruptly displacing the harmonic trap through a distance $\Delta
x$ along the $x$-direction \cite{X,Pasquini,preprint}. After the
displacement, the center of the trap coincides with the Gaussian
potential energy barrier and the total potential energy of the trap
and laser beam, for motion in the $x-y$ plane, is given by
\begin{eqnarray}
V(x,y)=m (\omega_x^2 x^2 +\omega_y^2 y^2)/2+V_L(x).
\end{eqnarray}
The solid curve in Fig 1(a) shows the form of $V(x,y=0)$. After the
displacement of the  harmonic trap, the condensate moves away from
its initial state [shown schematically by the dashed curve in Fig.
1(a)], reaches the Gaussian potential with a mean incident velocity
$\overline{v}_x \approx \omega_x \Delta x$, and is then scattered by
the potential barrier.
\begin{figure}
\includegraphics[width=7.0cm]{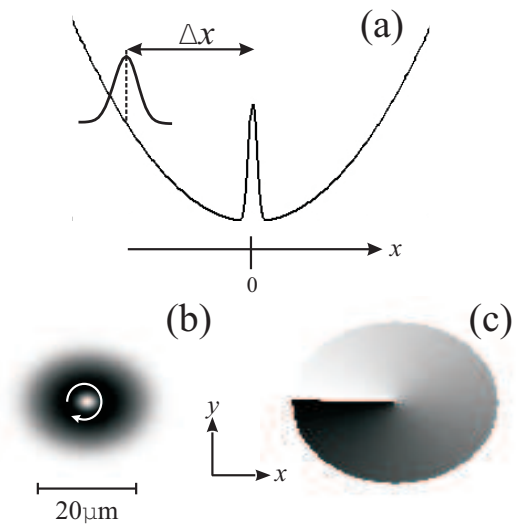}
\caption{(a) Solid curve: A schematic of the potential energy
$V\left(x,y=0\right)$ created by the harmonic  trap and laser beam
when $t  \ge 0$; Dashed curve:  a schematic of the initial
probability density $|\Psi(x,y=0,t=0)|^2$ for BEC $B$. (b) The
initial density $|\Psi(x,y,t=0)|^2$
 (black$=$high density, white$=$zero) of BEC $B$, where the white arrow shows the
 direction of the condensate's circulation and the horizontal bar
 denotes the scale. (c) The equivalent phase plot, $\phi(x,y,t=0)$,
 (white$=0$, black$=2\pi$).}
 \vspace{-0.5cm}
\end{figure}

The time-dependent Gross-Pitaevskii equation for the system is
\begin{eqnarray}
i\hbar  \frac{\partial \Psi \left(x,y,t\right)}{\partial
t}&=&-\frac{\hbar^2}{2m}\left[ \frac{\partial^2}{\partial
x^2}+\frac{\partial^2}{\partial y^2} \right]
\Psi\left(x,y,t\right) \nonumber \\ &+& \left[V\left(x,y\right)
 +
U_0 \mid  \Psi \left(x,y,t\right) \mid^2 \right]
\Psi \left(x,y,t\right) \nonumber \\
\end{eqnarray}
where  $ \Psi \left(x,y,t\right)$ is the wavefunction for motion in
the $x-y$ plane at time $t\ge 0$ and $U_0=4 \pi \hbar^2 a / m$ where
the $s$-wave scattering length  $a=2.9 \, $nm.

We  determine the initial BEC wave function by solving Eq. (3) for
$t \le 0$ using an imaginary time algorithm \cite{Imaginary}. When
the initial state contains a vortex, we impose the requirement that
there is a $2 \pi$-phase change in the condensate wavefunction
around the trap center at $(x,y)=(-\Delta x,0)$, which corresponds
to a quantized angular momentum of $-\hbar$ about the $z$-axis. The
wavefunction is normalized according to
\begin{equation}
\int \mid \Psi \left(x,y,t\right) \mid^2 dx dy=\frac{N}{L_z},
\end{equation}
where $L_z$ is the confinement length in the $z$-direction. For BECs
$A$, $B$ and $C$ we choose $N/L_z$ as $2.5 \times 10^8$ m$^{-1}$, $5
\times 10^9$ m$^{-1}$ and $2.5 \times 10^{11}$ m$^{-1}$
respectively,  which gives corresponding peak atom densities
$n_0=2.8 \times 10^{18}$ m$^{-3}$, $2.1 \times 10^{19}$ m$^{-3}$ and
$1.6 \times 10^{20}$ m$^{-3}$. The density, $\mid
\Psi(x,y,t=0)\mid^2$, and phase, $\phi(x,y,t=0)$, of the initial
state for  
BEC $B$ are shown in Figs. 1(b,c). Having obtained the initial state
of the BEC, we determine its motion by solving the Gross-Pitaevskii
equation (3) numerically using the Crank-Nicolson method
\cite{Crank}.

\section{NUMERICAL RESULTS FOR SCATTERING OFF A GAUSSIAN POTENTIAL BARRIER}
\begin{figure}
\includegraphics[width=8.5cm]{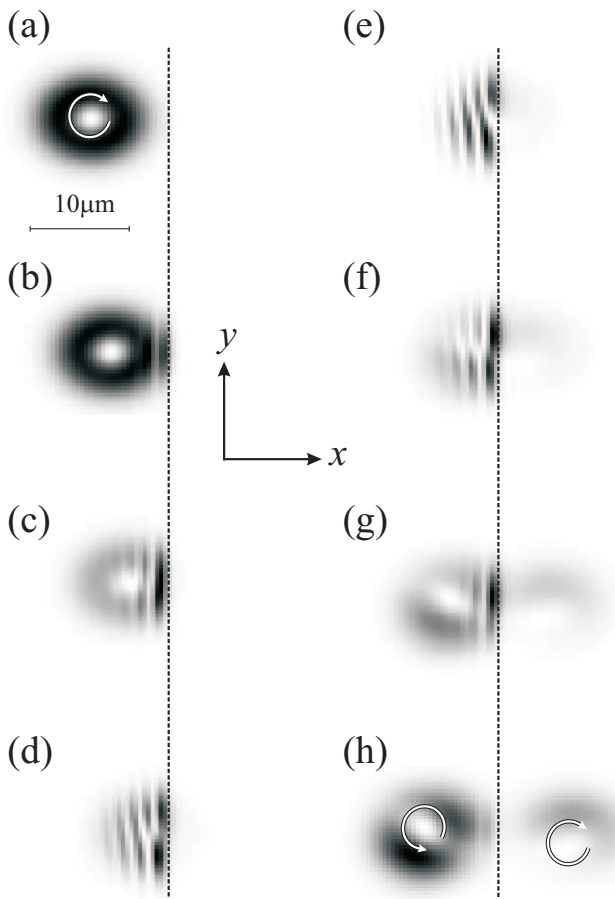}
\caption{Evolution of BEC $A$: plots of $|\Psi(x,y,t)|^2$
(black$=$high density, white$=$zero) for $\overline{v}_x =6.3 \,
$mm$\,$s$^{-1}$ at $t=$ $3.6\,$ms (a), $4.0\,$ms (b), $4.4\, $ms
(c), $4.8 \, $ms (d), $5.2 \,$ms (e), $5.6 \,$ms (f), $6.0 \,$ms (g)
and $6.4 \,$ms (h).  Coordinate axes are inset and the horizontal
bar indicates the scale. The arrows in (a) and (h) show the
direction of rotation.} \vspace{-0.5cm}
\end{figure}

In Fig. 2, we show the density profile of 
BEC $A$, with $\Delta x =20\, \mu$m ($\overline{v}_x =6.3
\, $mm$\,$s$^{-1}$) at $t=$ $3.6\,$ms (a), $4.0\,$ms (b), $4.4\, $ms
(c), $4.8 \, $ms (d), $5.2 \,$ms (e), $5.6 \,$ms (f), $6.0 \,$ms (g)
and $6.4 \,$ms (h). Figs. 2(a-e) show that as the BEC approaches the
Gaussian scattering potential at $x=0$ (dashed lines in Fig. 2), a
standing wave forms due to interference between the incoming and
reflected matter-waves. 
In Figs. 2(d,e) the standing wave undergoes a $\pi$ phase shift
between the upper (large $y$) and lower (small $y$) edges of the
BEC.
 This is due to the non-uniform initial phase of the
BEC, shown in Fig. 1(c).
After scattering, the BEC splits into reflected ($x<0$) and
transmitted ($x>0$) components [Figs. 2(f-h)]. Each of the
components contains a vortex  [enclosed by the arrows in Fig. 2(h)].
For the transmitted cloud, the quantized circulation is in the same
direction as in the incident cloud. By contrast, the reflected cloud
rotates in the opposite direction \cite{opposite1,opposite2}.
Physically, this is because atoms towards the top of the rotating
incident cloud, in Fig. 2, approach the barrier with a higher
velocity component along the $x$-direction ($v_x$) than those at the
bottom. But because $v_x$ is reversed after reflection, the
direction of rotation is also reversed in the reflected cloud. This
is in contrast to the transmitted atoms, which emerge through the
barrier with little change in their velocity and so the direction of
rotation is preserved. Since the trap remains switched on throughout
the scattering process, we do not observe the splitting of the
reflected component of the BEC that was reported in Ref.
\cite{opposite2}.
\begin{figure}
\includegraphics[width=8.5cm]{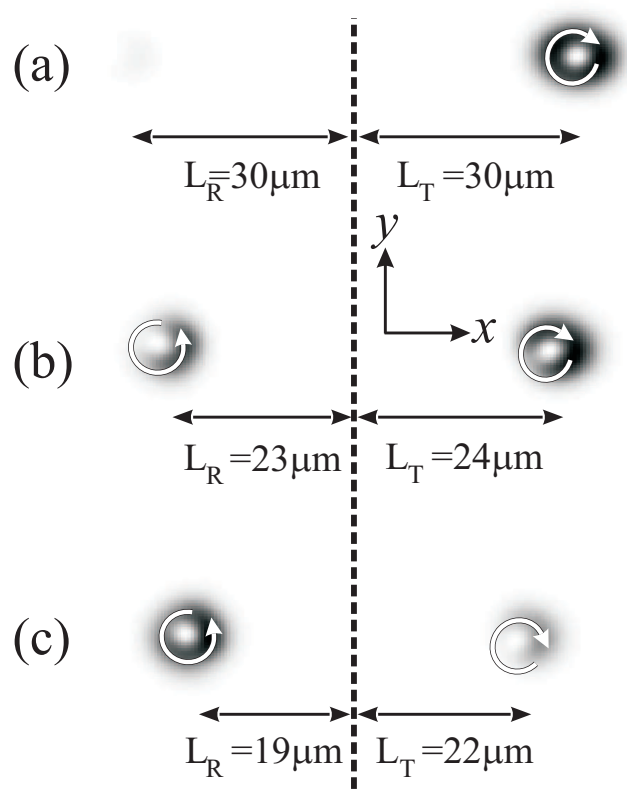}
\caption{BEC $A$: plots of $|\Psi(x,y,t=10\,$ms$)|^2$ for
$\overline{v}_x =9.4 \, $mm$\,$s$^{-1}$ (a), $7.5 \, $mm$\,$s$^{-1}$
(b) and $6.3 \, $mm$\,$s$^{-1}$ (c).  The circular arrows indicate
the direction of rotation and the horizontal arrows show the values
of $L_R$ and $L_T$, defined in the text and footnote
\cite{footnote}. Co-ordinate axes are inset. } \vspace{-0.5cm}
\end{figure}

We now consider how changing $\overline{v}_x$ affects the dynamics
of the atom cloud. Figure 3 shows the density profile of BEC $A$ at
$t=\pi/\omega_x=10 \,$ms after trap displacements of $30\, \mu$m
(a), $24 \, \mu$m (b) and $20 \, \mu$m (c), corresponding to
$\overline{v}_x=9.4\,$mm$\,$s$^{-1}$, $7.5\,$mm$\,$s$^{-1}$ and
$6.3\,$mm$\,$s$^{-1}$ respectively. Since this time is half the
period of oscillation of  the trap, both the transmitted and
reflected portions of the condensate will {\it approximately} be at
their turning points, and hence {\it nearly} stationary. In Fig. 3,
the vertical dashed line marks the trap center at $x=0$. When
$\overline{v}_x =9.4 \, $mm$\,$s$^{-1}$, the average kinetic energy
of the atoms incident upon the barrier is $10 \,$peV, which exceeds
the Gaussian barrier height. Hence, most of the condensate is
transmitted [Fig. 3(a)]. As we decrease $\overline{v}_x$ [Figs.
3(b,c)], the average energy of the incident atoms decreases and more
of the condensate is reflected. For $\overline{v}_x =6.3 \,
$mm$\,$s$^{-1}$ [Fig. 3(c)], most of the atoms are reflected by the
barrier because the average kinetic energy of the incident atoms is
$4.6 \,$peV, which is less than the barrier height. Since the BEC
has a finite spatial width along the $x$-direction of $l_x = 13
\,\mu$m, atoms that are towards the left-hand side of the BEC's
initial state travel further before reaching the barrier and
therefore have a higher incident velocity. Such atoms have a higher
transmission probability and therefore form a large fraction of the
transmitted atom cloud. We therefore expect that the distance
($L_T$) that the higher velocity transmitted cloud travels past the
scattering potential before coming to rest in the harmonic trap will
be greater than the distance ($L_R$) that the slower reflected cloud
retreats from the scatterer before reaching the turning point of the
harmonic trap \cite{footnote}. Our numerical simulations confirm
this: for example, in Fig. 3,  $L_T \ge L_R$ for all the values of
$\overline{v}_x$.
\begin{figure}
\includegraphics[width=8.0cm]{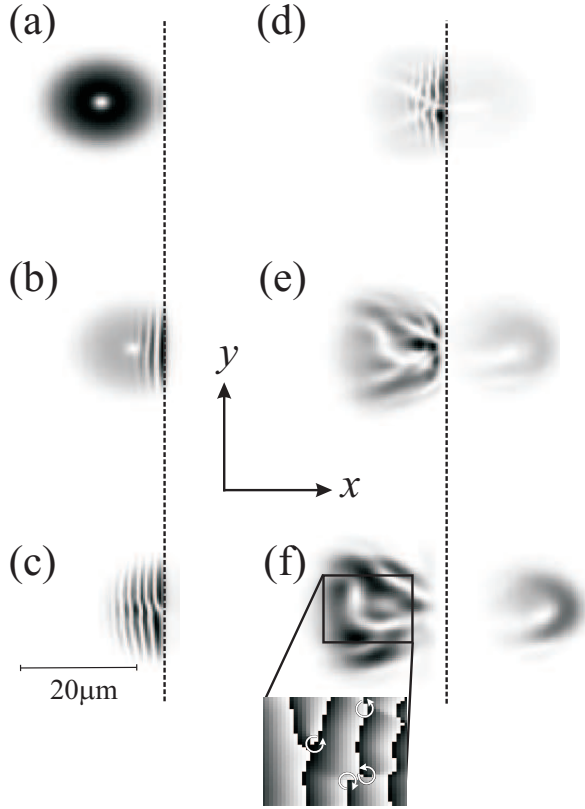}
\caption{Evolution of BEC  $B$: plots of $|\Psi(x,y,t)|^2$
(black$=$high density, white$=$zero) for $\overline{v}_x =6.3 \,
$mm$\,$s$^{-1}$ at $t=$ $3\,$ms (a), $4\,$ms (b), $5\, $ms (c), $6
\, $ms (d), $7 \,$ms (e), $8 \,$ms (f). The dashed line at $x=0$
marks the point where the laser potential is maximal. Co-ordinate
axes are inset and the horizontal bar indicates scale. Lower plot:
phase $\phi(x,y,t=8 \,$ms$)$ [white$=0$, black$=2\pi$] within the
region enclosed by the box in (f). Arrows indicate the direction of
circulation.} \vspace{-0.5cm}
\end{figure}

\begin{figure}
\includegraphics[width=8.5cm]{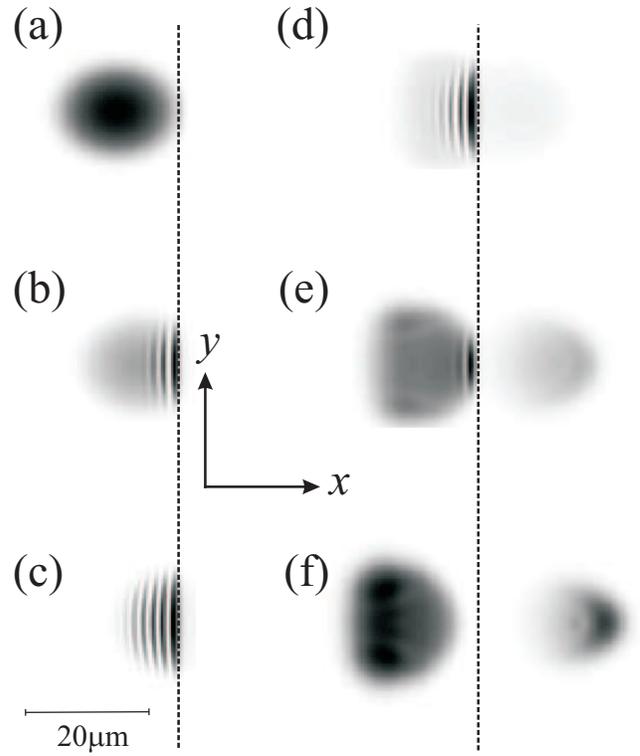}
\caption{Evolution of BEC $B_i$: plots of $|\Psi(x,y,t)|^2$
(black$=$high density, white$=$zero) for $\overline{v}_x =6.3 \,
$mm$\,$s$^{-1}$ at $t=$ $3\,$ms (a), $4\,$ms (b), $5\, $ms (c), $6
\, $ms (d), $7 \,$ms (e), $8 \,$ms (f). The dashed line at $x=0$
marks the point where the laser potential is maximal. Co-ordinate
axes are inset and the horizontal bar indicates scale.}
\vspace{-0.5cm}
\end{figure}

We now consider how the higher density BEC $B$ reflects off the
potential barrier. Figure 4 shows the density profile of  BEC $B$ at
$t=3\,$ms (a), $4 \,$ms (b), $5\,$ms (c), $6 \,$ms (d), $7\,$ms (e)
and $8 \,$ms (f), after a trap displacement of $20 \, \mu$m
($\overline{v}_x =6.3 \, $mm$\,$s$^{-1}$). As the BEC impinges upon
the Gaussian potential barrier, a standing wave forms between the
incoming and reflected matter waves. Figure 4(b) shows the first
stage of the standing wave formation in which maxima (black) and
nodal lines (white) appear at the leading edge of the atom cloud. In
contrast to BEC $A$,
 the reflected component of the BEC is significantly disrupted,
 as shown in  Figs. 4(e,f).
 This disruption is accompanied by the formation of {\it new}
vortices within the boxed region in Fig. 4(f).  The phase
variation, $\phi(x,y,t=8 \, $ms$)$, within this region is shown in
the lower part of Fig. 4(f). Around each vortex, $\phi$ increases
from $0$ to $2\pi$ in the direction of circulation shown by the
arrows in Fig. 4(f).

To explain the disruption shown in Fig. 4,  we first consider the
dynamics of BEC $B_i$. Figure 5 shows density profiles for this
initially irrotational condensate at $t=3\,$ms (a), $4 \,$ms (b),
$5\,$ms (c), $6 \,$ms (d), $7\,$ms (e) and $8 \,$ms (f) after a trap
displacement $20 \, \mu$m ($\overline{v}_x =6.3 \, $mm$\,$s$^{-1}$).
No dynamical excitations are produced in the transmitted or
reflected clouds. We therefore conclude that the disruption observed
for BEC $B$ (Fig. 4) is related to the rotation of
the cloud. 
\begin{figure}
\includegraphics[width=8.5cm]{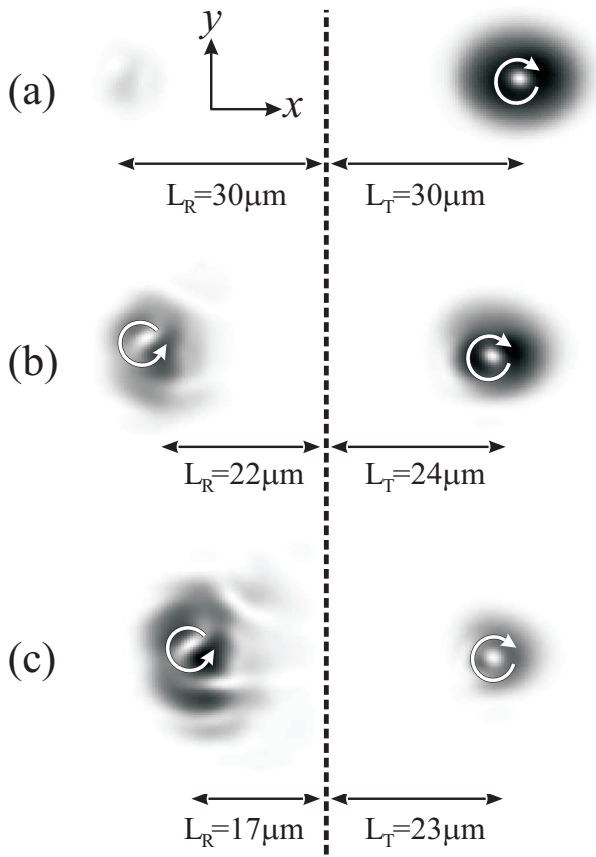}
\caption{BEC $B$: plots of $|\Psi(x,y,t=10\,$ms$)|^2$ for
$\overline{v}_x =9.4 \, $mm$\,$s$^{-1}$ (a), $\overline{v}_x =7.5 \,
$mm$\,$s$^{-1}$ (b) and $\overline{v}_x =6.3 \, $mm$\,$s$^{-1}$ (c).
The dashed line at $x=0$ marks the point where the laser potential
is maximal. The circular arrows indicate the direction  of rotation
and the horizontal arrows show the values of $L_R$ and $L_T$,
defined in the text and footnote \cite{footnote}. Co-ordinate axes
are inset.} \vspace{-0.5cm}
\end{figure}


To gain further insights into this disruption,  we now examine how
changing $\overline{v}_x $ affects the transmission and reflection
of BEC $B$. Figure 6 shows the condensate density at $t=10 \, $ms
after trap displacements of $\Delta x= 30 \, \mu$m (a), $24 \, \mu$m
(b), and $20 \, \mu$m (c), corresponding to mean incident velocities
of $\overline{v}_x=9.4\,$mm$\,$s$^{-1}$, $7.5\,$mm$\,$s$^{-1}$ and
$6.3\,$mm$\,$s$^{-1}$ respectively. For $\overline{v}_x =9.4 \,
$mm$\,$s$^{-1}$, the behavior of BEC $B$ [Fig. 6(a)] is
qualitatively the same as BEC $A$ [Fig. 3(a)]: the transmitted cloud
retains the vortex structure of the incident cloud and only a small
fraction of atoms are reflected. As $\overline{v}_x$ decreases
[Figs. 6(b,c)], the reflection probability rises, but the structure
of the reflected cloud becomes increasingly disrupted. This
disruption contrasts with BEC $A$, where for all $\overline{v}_x$
values the reflected cloud retains a well defined vortex structure
(Fig. 3). For BEC $B$, we find that $L_T \ge L_R$ (see values in
Fig. 6) just as for BEC $A$. However, for $\overline{v}_x =7.5 \,
$mm$\,$s$^{-1}$ and $\overline{v}_x =6.3 \, $mm$\,$s$^{-1}$, $L_R$
is lower for BEC $B$ [Fig. 6(b,c)] than for BEC $A$ [Fig. 3(b,c)]
due to the fragmentation of the atom cloud that occurs for BEC $B$
only. This fragmentation transfers kinetic energy from the
center-of-mass motion of the reflected cloud into internal
vorticity, thus damping the oscillation.

The effect of the vortex on the scattering process depends on the
scattering time of the vortex core ($t_{sv}$) relative to its
rotation time ($t_r$). Since the diameter of the vortex core is
approximately the healing length $\xi=1/\sqrt{8 \pi n_0 a}$, it
follows that $t_{sv}\approx 2 \xi / \overline{v}_x$ and $t_r=\pi
\xi^2 m/h$. If the ratio
\begin{eqnarray}
\frac{t_r}{t_{sv}}=\frac{m}{2h} \pi \xi \Delta x \omega_x \propto
\frac{\Delta x}{\sqrt{n_0}}
\end{eqnarray}
is $\gg 1$, so that there is insufficient time for the vortex to
rotate during  the scattering process,  the vortex will have little
effect on the dynamics of the transmitted and reflected atom clouds.
Physically, this is because all parts of the BEC are incident on the
Gaussian potential with a variation in the incident velocity, due to
the rotation, that is small compared to $\overline{v}_x$.
Conversely, if $t_r / t_{sv} \lappeq 1$ we expect that the rotation
of the BEC will disrupt the reflected component of the BEC. We refer
to such a regime as rotational disruption. For the lower-density BEC
$A$, $t_r/t_{sv} \gg 1$ for all three of the $\overline{v}_x$ values
considered in Fig. 3, and so the reflected atom cloud is not
significantly fragmented. 
By contrast, for the higher density BEC $B$ $t_r/t_{sv}<1$, and
hence rotational disruption occurs in the reflected atom cloud. Note
that reducing $\overline{v}_x$ from $9.4 \, $mm$\,$s$^{-1}$ [Fig.
6(a)] to $6.3 \, $mm$\,$s$^{-1}$ [Fig. 6(c)] causes $t_r/t_{sv}$ to
decrease, and thus increases the  fragmentation of the reflected
cloud.

\begin{figure}
\includegraphics[width=8.5cm]{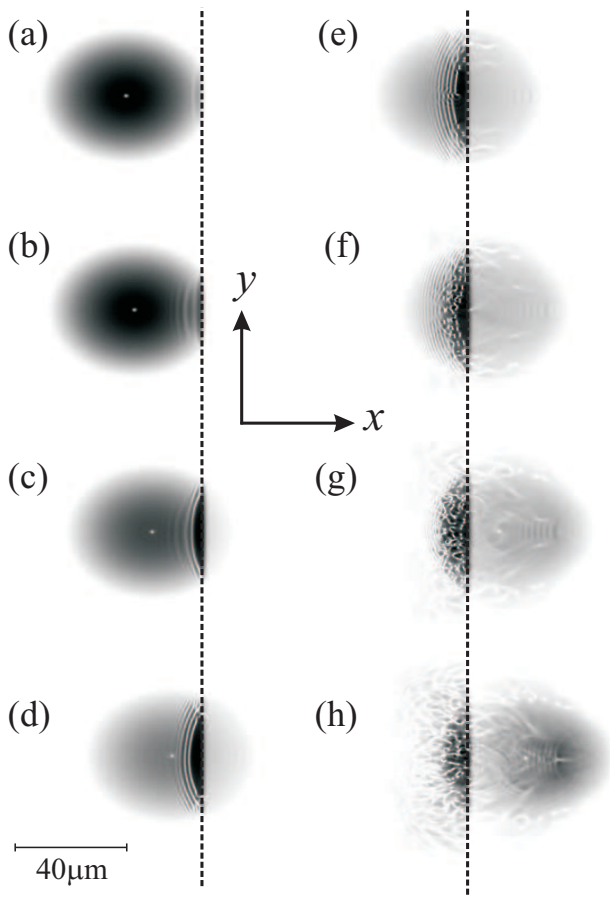}
\caption{Evolution of BEC $C$: plots of $|\Psi(x,y,t)|^2$
(black$=$high density, white$=$zero) for $\overline{v}_x =9.4 \,
$mm$\,$s$^{-1}$ at $t=$ $1\,$ms (a), $2\,$ms (b), $3\, $ms (c), $4
\, $ms (d), $5 \,$ms (e), $6 \,$ms (f),  $7 \, $ms (g) and $8 \,$ms
(h). Co-ordinate axes are inset and the horizontal bar indicates the
scale.} \vspace{-0.5cm}
\end{figure}
\begin{figure}
\includegraphics[width=8.5cm]{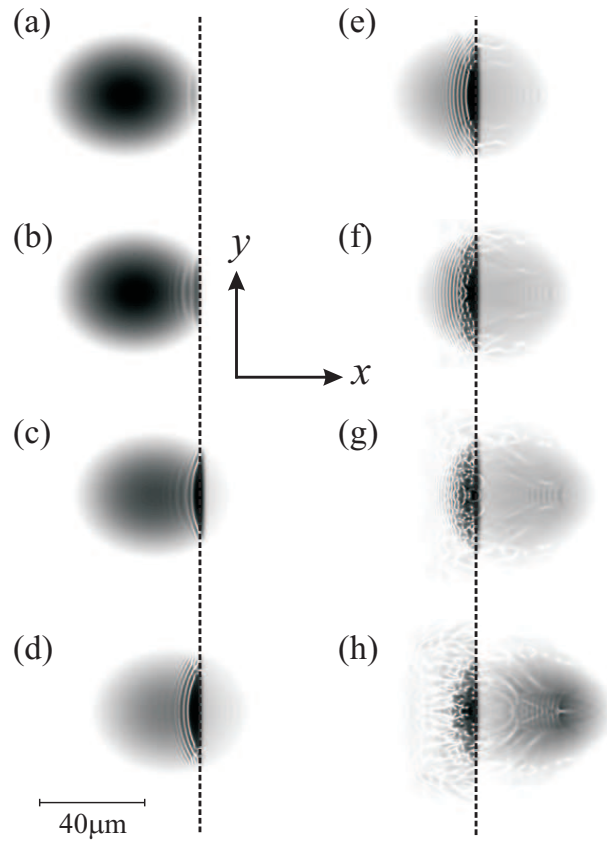}
\caption{Evolution of BEC $C_i$: plots of $|\Psi(x,y,t)|^2$
(black$=$high density, white$=$zero) for $\overline{v}_x =9.4 \,
$mm$\,$s$^{-1}$ at $t=$ $1\,$ms (a), $2\,$ms (b), $3\, $ms (c), $4
\, $ms (d), $5 \,$ms (e), $6 \,$ms (f),  $7 \, $ms (g) and  $8 \,$ms
(h). Co-ordinate axes are inset and the horizontal bar indicates the
scale.} \vspace{-0.6cm}
\end{figure}
Disruption of the reflected cloud is even more pronounced for the
higher density BEC $C$. Figure 7 shows the density of this BEC at
$t=$ $1\,$ms (a), $2\,$ms (b), $3\, $ms (c), $4 \, $ms (d), $5 \,$ms
(e), $6 \,$ms (f),  $7 \, $ms (g) and $8 \,$ms (h) after a trap
displacement of $30 \,\mu$m ($\overline{v}_x =9.4 \,
$mm$\,$s$^{-1}$) \cite{note}. As the condensate impinges upon the
scattering potential, a standing wave forms between the incident and
reflected matter waves, Figs. 7(b-d). This standing wave seeds
solitons \cite{PRL,PRA,preprint}, which decay via the snake
instability \cite{snake} into vortex-antivortex pairs [Figs.
7(e,f)], thus strongly disrupting the internal structure of the
cloud [Figs. 7(g,h)]. Consequently, when incident atoms subsequently
pass through the barrier they produce some irregularity in the
transmitted atom cloud. Note, that this irregularity is less
pronounced towards the right hand edge of the transmitted cloud,
which contains the atoms that passed through the barrier {\it
before} solitons and vortices formed at negative $x$ ($x=0$ is
marked by the dotted line in Fig. 7). We emphasize that although BEC
$C$ is in a regime where rotational disruption occurs, the initial
vortex does {\it not} (in contrast to BEC $B$) cause the severe
fragmentation shown in Fig. 7. To demonstrate this, Fig. 8 shows
 BEC $C_i$ (identical to BEC $C$, except irrotational) scattering
off a Gaussian potential at $t=$ $1\,$ms (a), $2\,$ms (b), $3\, $ms
(c), $4 \, $ms (d), $5 \,$ms (e), $6 \,$ms (f),  $7 \, $ms (g) and
$8 \,$ms (h) after a trap displacement of $30 \, \mu$m
($\overline{v}_x =9.4 \, $mm$\,$s$^{-1}$). Comparison of Figs. 7 and
8 reveals that the dynamics are qualitatively the same irrespective
of whether or not the BEC contains an initial vortex.

To understand these results, we recall previous work
\cite{PRA1,PRL,PRA,Y} on the Bragg  reflection of a BEC in an
optical lattice. In Refs. \cite{PRL,PRA} it was shown that at Bragg
reflection, fragmentation can arise from the density and phase
imprinting that accompanies standing wave formation. When the
correlation time,
\begin{eqnarray}
t_c = \frac{m}{2\sqrt{2}hn_0a},
\end{eqnarray}
is much less than the Bloch period ($t_B$), this imprinting leads to
the formation of solitons and vortices,  which disrupt the atom
cloud. For BECs $C$ and $C_i$, a similar disruption occurs when $t_c
\ll t_s$, where $t_s=l_x/\overline{v}_x$ is the approximate duration
of the reflection process.  This effect is described as {\it
interferential disruption} \cite{Robin}, since it originates from
the interference pattern, in this case produced by the superposition
of the incident and reflected matter waves. For BEC $C$ this
interferential disruption dominates the dynamics and completely
masks any effects due to the rotational disruption.


\section{REFLECTION PROBABILITY OF THE BEC}
In this section, we investigate how the reflection probability of
the BEC varies with $\overline{v}_x$ and depends on the presence or
absence of an initial vortex. Defining the reflection probability of
the condensate as
\begin{equation}
R(\overline{v}_x=\Delta x \omega_x)=\frac{\int_y \int_{x <0} \mid
\Psi  \left(x,y,t=\pi/\omega_x \right) \mid^2 dx dy}{\int_y \int_x
\mid \Psi \left(x,y,t\right) \mid^2 dx dy},
\end{equation}
we use our numerical solution of the Gross-Pitaevskii equation to
quantify the reflection  probabilities of BEC $A$ [$R
(\overline{v}_x)$, dotted curve in Fig. 9(a)] and its irrotational
equivalent [$R_{i}(\overline{v}_x)$, dashed curve in Fig. 9(a)].
Figure 9(a) shows that as $\overline{v}_x$ decreases, both
$R(\overline{v}_x)$ and $R_{i}(\overline{v}_x)$ increase. For
$\overline{v}_x \lappeq 7.3\, $mm s$^{-1}$, $R(\overline{v}_x) <
R_i(\overline{v}_x)$ and conversely for $\overline{v}_x \gappeq
7.3\, $mm s$^{-1}$, $R(\overline{v}_x) > R_{i}(\overline{v}_x)$.
This crossover is revealed more clearly by the dashed curve in Fig.
9(b), which shows $r(\overline{v}_x)=R(\overline{v}_x) -
R_{i}(\overline{v}_x)$. For $\overline{v}_x \lappeq 7.3 \, $mm
s$^{-1}$, $r(\overline{v}_x)< 0$, but at higher $\overline{v}_x$,
$r(\overline{v}_x)>0$.
\begin{figure}
\includegraphics[width=5.5cm]{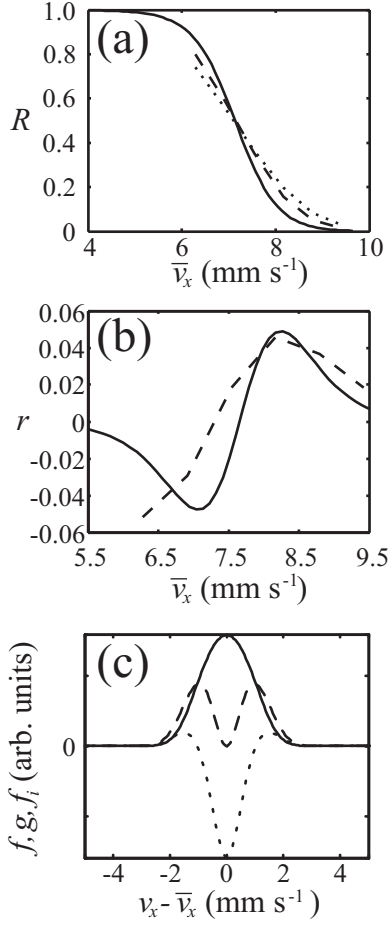}
\caption{(a) Dotted curve: reflection probability,
$R(\overline{v}_x)$, for BEC $A$. Dashed curve: reflection
probability, $R_{i}(\overline{v}_x)$, for BEC $A_i$. Solid curve:
reflection probability, $R_{P}(\overline{v}_x)$, for a plane wave.
(b)  Dashed curve: $r(\overline{v}_x)$ obtained from numerical
simulations. Solid curve: the expression for $r(\overline{v}_x)$
obtained from Eq. (18) with $I=0.007 \,$mm$^2$ s$^{-2}$ and $J=0.19
\,$mm$^2$ s$^{-2}$. (c) Solid curve: the local velocity distribution
function, $f_i(v_x-\overline{v}_x, y=0)$ for BEC $A_i$. Dashed
curve: the local velocity distribution function,
$f(v_x-\overline{v}_x,y=0)$ for BEC $A$. Dotted curve:
$g(v_x-\overline{v}_x,y=0)$.} \vspace{-0.5cm}
\end{figure}

In Fig. 9(a) we compare $R(\overline{v}_x)$ and
$R_{i}(\overline{v}_x)$ to the equivalent plane wave reflection
probability $R_P(\overline{v}_x)$ (solid curve). The difference
between $R_{P}(\overline{v}_x)$ and $R_{i}(\overline{v}_x)$ is due
to the finite width of the BEC along the $x$-direction, which means
that the atoms start from various positions in the harmonic trap and
thus reach the barrier with a range of incident velocities rather
than the single velocity of an incident plane wave. We can model the
spread of velocities by defining the probability
$F_i(v_x-\overline{v}_x)dv_x=\int_y f_i(v_x-\overline{v}_x,y)dydv_x$
that an atom arrives at the barrier with an incident velocity
between $v_x$ and $v_x+dv_x$. Here
$f_i(v_x-\overline{v}_x,y)dv_x=|\Psi((v_x-\overline{v}_x)/\omega_x,y,t=0)|^2dv_x/\int_{v_x}
|\Psi((v_x-\overline{v}_x)/\omega_x,y,t=0)|^2 dv_x$ is the {\it
local} probability that an atom starting from position $y$ arrives
at the barrier with a normal velocity component between $v_x$ and
$v_x+dv_x$. For each $y$, this local probability is determined from
the initial shape of the BEC wavefunction, which defines the initial
spatial distribution of the atoms and hence the resulting velocity
distribution as the atoms arrive at the barrier. For each $y$,
$f_i(v_x-\overline{v}_x,y)$ is symmetrical about
$v_x-\overline{v}_x=0$ owing to the symmetry of the initial
wavefunction. This can be seen from the solid curve in Fig. 9(c),
which shows $f_i(v_x-\overline{v}_x,y=0)$ for BEC $A_i$.
The reflection probability of the BEC can be expressed in terms of
its velocity distribution and the reflection probability of a plane
wave by
\begin{eqnarray}
R_{i}(\overline{v}_x) = \int_{0}^{\infty} F_i(v_x
-\overline{v}_x)R_{P}(v_x)dv_x.
\end{eqnarray}
Since $F_i(v_x-\overline{v}_x)$  is sharply peaked at
$v_x=\overline{v}_x$ we can use the Taylor expansion
\begin{eqnarray}
R_{P}(v_x) &\approx& R_{P}  (\overline{v}_x)+
\left. \frac{\partial R_{P}(v_x)}{\partial v_x}\right|_{\overline{v}_x}\left(v_x-\overline{v}_x \right) \nonumber \\
&+&\left. \frac{\partial^2 R_{P}(v_x)}{2\partial v_x^2}
\right|_{\overline{v}_x} \left( v_x - \overline{v}_x \right)^2
\end{eqnarray}
to write Eq. (8) in the form
\begin{eqnarray}
R_{i}(\overline{v}_x) &\approx&
R_{P}(\overline{v}_x)\int_{0}^{\infty} F_i(v_x -\overline{v}_x)dv_x
\nonumber \\ &+&
\left. \frac{\partial R_{P}(v_x)}{\partial v_x}\right|_{\overline{v}_x}\int_{0}^{\infty} F_i(v_x -\overline{v}_x)(v_x-\overline{v}_x)dv_x \nonumber \\
&+&\left. \frac{\partial^2 R_{P}(v_x)}{2\partial
v_x^2}\right|_{\overline{v}_x} \int_{0}^{\infty} F_i(v_x
-\overline{v}_x)(v_x-\overline{v}_x)^2dv_x .\nonumber \\ & &
\end{eqnarray}
Since  $F_i(v_x -\overline{v}_x)$ is symmetrical about $0$  and its
width is less than $\overline{v}_x$ for all the $\overline{v}_x$
values considered in Figs. 9(a,b), Eq. (10) reduces to
\begin{eqnarray}
R_{i}(\overline{v}_x) \approx R_{P}(\overline{v}_x) +
\frac{I}{2}\left. \frac{\partial^2 R_{P}(v_x)}{\partial
v_x^2}\right|_{\overline{v}_x} ,
\end{eqnarray}
where \cite{notef2}
\begin{eqnarray}
I=\int_{0}^{\infty} F_i(v_x
-\overline{v}_x)(v_x-\overline{v}_x)^2dv_x
>0.
\end{eqnarray}
It follows from Eqs. (11) and (12) that $R_{i}(\overline{v}_x) >
R_P(\overline{v}_x)$ if
\begin{eqnarray}
\left. \frac{\partial^2 R_{P}(v_x)}{\partial v_x^2}\right|_{\overline{v}_x}>0
\end{eqnarray}
and  $R_{i}(\overline{v}_x) < R_{P}(\overline{v}_x)$ if
\begin{eqnarray}
\left. \frac{\partial^2 R_{P}(v_x)}{\partial v_x^2}\right|_{\overline{v}_x}<0.
\end{eqnarray}
Evaluating the integral in Eq. (12) numerically, we find that for
BEC $A_i$, $I=0.007\, $mm$^2$ s$^{-2}$.

From the shape of the solid curve in Fig. 9(a), we see that for
$\overline{v}_x \lappeq 7.3\, $mm s$^{-1}$, $ \left.
\frac{\partial^2 R_{P}(v_x)}{\partial v_x^2}\right|_{\overline{v}_x}
< 0$, whilst for $\overline{v}_x \gappeq 7.3\,$mm s$^{-1}$, $ \left.
\frac{\partial^2 R_{P}(v_x)}{\partial v_x^2}\right|_{\overline{v}_x}
> 0$. Hence, from Eq. (11), we expect $R_{i}(\overline{v}_x)
>R_{P}(\overline{v}_x)$, for  $\overline{v}_x \gappeq 7.3 \,$mm
s$^{-1}$, and $R_{i}(\overline{v}_x) <R_{P}(\overline{v}_x)$, for
$\overline{v}_x \lappeq 7.3 \,$mm s$^{-1}$, which is confirmed by
comparing the dashed [$R_{i}(\overline{v}_x)$] and solid
[$R_{P}(\overline{v}_x)$] curves in Fig. 9(a).

We now consider a BEC that initially contains a single vortex. In
this case, Eq. (8) can be generalized to
\begin{eqnarray}
R(\overline{v}_x) = \int_{0}^{\infty} F(v_x
-\overline{v}_x)R_{P}(v_x)dv_x
\end{eqnarray}
where the velocity distribution for the incident atoms in the
rotating BEC, $F(v_x-\overline{v}_x)$, equals the velocity
distribution function for an irrotational BEC plus a correction
$G(v_x -\overline{v}_x)$ that is
\begin{eqnarray}
F(v_x-\overline{v}_x)=F_i(v_x-\overline{v}_x)+G(v_x
-\overline{v}_x),
\end{eqnarray}
where $F(v_x-\overline{v}_x)=\int_y f(v_x-\overline{v}_x,y)dy$ and
$G(v_x-\overline{v}_x)=\int_y g(v_x-\overline{v}_x,y)dy$ are defined
in terms of the local velocity distribution functions
$f(v_x-\overline{v}_x,y)$ and
$g(v_x-\overline{v}_x,y)=f(v_x-\overline{v}_x,y)-f_i(v_x-\overline{v}_x,y)$,
which are shown in Fig. 9(c), for $y=0$, by the dashed and dotted
curves respectively. Since $g(v_x-\overline{v}_x,y)$ and
$G(v_x-\overline{v}_x)$ are both symmetrical about
$v_x-\overline{v}_x=0$, it follows from an analysis similar to that
presented in Eqs. (8-10) that
\begin{eqnarray}
R(\overline{v}_x) &\approx& R_{i}(\overline{v}_x)  \nonumber \\
&+&\left. \frac{\partial^2 R_{P}(v_x)}{2\partial v_x^2}\right|_{\overline{v}_x} \int_{0}^{\infty} G(v_x -\overline{v}_x)(v_x-\overline{v}_x)^2dv_x, \nonumber \\
& &
\end{eqnarray}
which can be rewritten as
\begin{eqnarray}
R(\overline{v}_x) &\approx& R_{i}(\overline{v}_x)  +\frac{J}{2}\left. \frac{\partial^2 R_{P}(v_x)}{\partial v_x^2}\right|_{\overline{v}_x} \nonumber \\
& \approx& R_P(\overline{v}_x) +\frac{\left(I+J\right)}{2} \left. \frac{\partial^2 R_{P}(v_x)}{\partial v_x^2}\right|_{\overline{v}_x}
\end{eqnarray}
where
\begin{eqnarray}
J=\int_{0}^{\infty} G(v_x
-\overline{v}_x)(v_x-\overline{v}_x)^2dv_x.
\end{eqnarray}

Unlike $I$, the sign of $J$ can be either negative or positive,
depending on the form of $G(v_x-\overline{v}_x)$. To highlight this,
we initially consider a simple analysis based on the form of the
local distribution function $g(v_x-\overline{v}_x,y=0)$ [Fig. 9(c)].
For $v_x \approx \overline{v}_x$ (i.e. in the region of the vortex
core) $f(v_x-\overline{v}_x,y=0)$ [dashed curve in Fig. 9(c)] is
close to zero, because the atom density falls to zero at the vortex
core, whereas $f_i(v_x-\overline{v}_x,y=0)$ [solid curve in Fig.
(c)] is maximal. Consequently, $g(v_x-\overline{v}_x,y=0)$ [dotted
curve in Fig. 9(c)] is negative for $v_x \approx \overline{v}_x$.
Since the atoms have moved away from the vortex core, towards the
edges of the BEC, we expect that away from the vortex core (where
$|v_x-\overline{v}_x| \gg 0$) $f(v_x-\overline{v}_x,y=0)
> f_i(v_x-\overline{v}_x,y=0)$ and hence $g(v_x-\overline{v}_x,y=0) > 0$, as
can be seen in Fig. 9(c) (dotted curve). Consequently, the
contribution to $J$ is likely to be positive because the integrand
in Eq. (19) is largest when $(v_x-\overline{v}_x)^2$ is large.

However, if $y \ne 0$, $v_x$ is also perturbed by the circulation of
the BEC around the vortex core. For $y>0$, $v_x$ is increased and
for $y<0$, $v_x$ is decreased. This effect further increases the
spread of incident velocities at the barrier and, due to the
$(v_x-\overline{v}_x)^2$ term in Eq. (19), makes $J$ more positive.
From our full numerical analysis we find that the dominant cause of
the shift in the reflection probabilities is the circulation of
atoms around the vortex. Evaluating Eq. (19) numerically, we find
for BEC $A$ that $J=0.19 \,$mm$^2$ s$^{-2} \gg I$.

Since $J$ and $I$  are both positive for the BEC considered here,
from Eqs. (11) and (18), we expect that
$R(\overline{v}_x)-R_P(\overline{v}_x)$ will have the same sign as,
but a larger magnitude than,
$R_{i}(\overline{v}_x)-R_P(\overline{v}_x)$. This is confirmed by
the curves shown in Fig. 9(a) and highlights the fact that the
presence of the vortex increases the distribution of incident
velocities upon the scattering potential.

In Fig. 9(b), we compare  the expression for $r(\overline{v}_x)
\approx \left.\frac{J}{2} \frac{\partial^2 R_P(v_x)}{\partial
v_x^2}\right|_{\overline{v}_x}$, obtained from Eq. (18) (solid
curve),  with the numerical values of $r(\overline{v}_x)$ (dashed
curve) obtained from our solutions of the Gross-Pitaevskii equation.
The expression for $r(\overline{v}_x)$ (solid curve) and the
numerical simulations (dashed curve) are in good agreement because
the spread of incident velocities is narrow and hence the Taylor
expansions used in the derivation of Eqs. (11) and (18) are
reasonably accurate.  For BEC $B$ we find that the discrepancy
between the numerical $r(\overline{v}_x)$ values and those obtained
from Eq. (18) becomes larger. This discrepancy is not only due to an
increase in the spread of incident velocities, which introduces
higher order terms into the Taylor expansion, but also arises from
the disruption of the BEC upon scattering, an effect which is not
described in our approximate analysis. Equations. (11) and (18) are
therefore only valid when no interferential or rotational disruption
is present in the BEC.

\section{CONCLUSIONS}
We have investigated how  BECs with different atom densities scatter
off a Gaussian potential when the initial state is either
irrotational or contains a single vortex.
We find three distinct regimes for the formation of dynamical
excitations and hence rotational or interferential disruption: (i)
at low densities there is no fragmentation of the reflected or
transmitted components of the BEC, irrespective of whether or not
there is a vortex in the initial state; (ii) at medium densities for
which $t_r/t_{sr} \lappeq 1$, rotational disruption occurs in the
reflected component of a BEC with an initial vortex, but no
disruption is observed when the BEC is initially irrotational; (iii)
at high densities, there is strong interferential disruption in the
reflected atom cloud if $t_c \lappeq t_s$, both in the presence and
absence of an initial vortex.

By considering the velocity distribution of the incident atoms, we
have derived expressions for the reflection probabilities of
rotating and irrotational BECs in terms of the reflection
probability of a single plane wave incident on the scattering
potential. This analytic approach agrees well with our numerical
calculations of reflection probabilities for BECs scattering from a
Gaussian barrier. It shows that the velocity spread of an
irrotational BEC causes a positive or negative deviation from
$R_P(\overline{v}_x)$, depending on the curvature of
$R_P(\overline{v}_x)$. When a vortex is introduced, the circulation
further increases the spread of incident velocities, leading to an
even larger deviation from $R_P(\overline{v}_x)$.

Finally, we note that with  current techniques it should be possible
to perform experimental tests of our theoretical predictions,
relating to the existence of three distinct dynamical regimes and to
the effect of a vortex on the reflection probability of a BEC
impinging on a potential barrier.

This work was supported by  the ARC, the EPSRC, the Royal Society
(London) and the University of Melbourne.


\begin{thebibliography}{99}
\bibitem{Morsch}O. Morsch, J.H. M\"uller, M. Cristiani and E.
Arimondo, Phys. Rev. Lett. {\bf 87}, 140402 (2001).
\bibitem{mirror1}A. S. Arnold, C. MacCormick, and M. G. Boshier,
Phys. Rev. A {\bf 65}, 031601 (2002).
\bibitem{mirror2}K. Bongs, S. Burger, G. Birkl, K. Sengstock, W. Ertmer, K. Rzazewski,
A. Sanpera, and M. Lewenstein, Phys. Rev. Lett. {\bf 83}, 3577
(1999).
\bibitem{X} A. G\"unther, S. Kraft, M. Kemmler, D. Koelle, R. Kleiner,
C. Zimmermann and J. Fort\'agh, Phys. Rev. Lett. {\bf 95}, 170405
(2005).
\bibitem{Pasquini} T.A. Pasquini, Y. Shin, C. Sanner, M. Saba, A. Schirotzek, D.E. Pritchard, and W.
Ketterle, Phys. Rev. Lett. {\bf 93}, 223201 (2004); T.A. Pasquini,
M. Saba, G.-B. Jo, Y. Shin, W. Ketterle, D.E. Pritchard, T.A. Savas
and N. Mulders, Phys. Rev. Lett. {\bf 97}, 093201 (2006).
\bibitem{preprint} R.G. Scott, A.M. Martin, T.M. Fromhold and F.W. Sheard, Phys. Rev. Lett. {\bf 95}, 073201 (2005).
\bibitem{vortex1}M. R. Matthews, B. P. Anderson, P. C. Haljan, D. S. Hall, C. E. Wieman, and E. A.
Cornell, Phys. Rev. Lett. {\bf 83}, 2498 (1999).
\bibitem{vortex2}K. W. Madison, F. Chevy, W. Wohlleben, and J.
Dalibard, Phys. Rev. Lett. {\bf 84}, 806 (2000).
\bibitem{vortex3}F. Chevy, K. W. Madison, and J. Dalibard, Phys.
Rev. Lett. {\bf 85}, 2223 (2000).
\bibitem{vortex4} C. Raman, J. R. Abo-Shaeer, J. M. Vogels, K. Xu, and W.
Ketterle, Phys. Rev. Lett. {\bf 87}, 210402 (2001).
\bibitem{PRA1} R.G. Scott, A.M. Martin and T.M. Fromhold, Phys. Rev. A {\bf 69}, 063607 (2004).
\bibitem{opposite1} J.J. García-Ripoll, G. Molina-Terriza, V. M. Pérez-García and L.
Torner, Phys. Rev. Lett. {\bf 87}, 140403 (2001).
\bibitem{opposite2}I. Josopait, L. Dobrek, L. Santos, A. Sanpera and M.
Lewenstein, Eur. Phys. J. D. {\bf 22}, 385 (2003).
\bibitem{Ketterle} Y. Shin, M. Saba, T.A. Pasquini, W. Ketterle, D.E. Pritchard and A.E. Leanhardt, Phys. Rev. Lett. {\bf 92}, 050405 (2004).
\bibitem{MOT} M. Greiner, O. Mandel, T.W. H${\rm \ddot{a}}$nsch and I. Bloch, Nature {\bf 419}, 51 (2002).
\bibitem{Robin}R.G Scott, D.A.W. Hutchinson and C.W. Gardiner,
cond-mat/0608135 (unpublished).
\bibitem{Imaginary}M.L. Chiofalo, S. Succi and M.P. Tosi, Phys.
Rev. E, {\bf 62}, 7438 (2000).
\bibitem{Crank}W.H. Press, S.A. Teukolsky, W.T. Vetterling and
B.P. Flannery, {\it Numerical Recipes, the Art of Scientific Computing}
(Cambridge University Press, Cambridge, 1994).
\bibitem{footnote}We calculate the expectation values of the positions of the transmitted and reflected  atom clouds when they come to rest at $t=\pi/\omega_x$ as
\begin{eqnarray}
L_T=\frac{\int_y \int_{x >0}  \mid \Psi \left(x,y;t=\pi/\omega_x \right) \mid^2x dx
dy}{\int_y \int_x \mid \Psi \left(x,y;t\right) \mid^2 dx dy}  \nonumber
\end{eqnarray}
and
\begin{eqnarray}
L_R=\frac{\int_y \int_{x <0}  \mid \Psi \left(x,y;t=\pi/\omega_x \right)\mid^2 x dx
dy}{\int_y \int_x \mid \Psi \left(x,y;t\right) \mid^2 dx dy}. \nonumber
\end{eqnarray}
\bibitem{note} Since the interactions in BEC $C$ play a dominant role, the diameter of the condensate is $50 \mu$m and hence it is only possible to consider $\Delta x > 25 \mu$m. For this reason we only consider $\Delta x =30 \mu$m for system $C$.
\bibitem{PRL} R.G. Scott, A.M. Martin, T.M. Fromhold, S. Bujkiewicz, F.W. Sheard and M.
Leadbeater, Phys. Rev. Lett. {\bf 90}, 110404 (2003).
\bibitem{PRA} R.G. Scott, A.M. Martin, S. Bujkiewicz, T.M. Fromhold,
N. Malossi, O. Morsch, M. Cristiani and E. Arimondo, Phys. Rev. A.
{\bf 69}, 033605 (2004).
\bibitem{Y} L. Fallani, L. De Sarlo, J.E. Lye, M. Modugno, R. Saers,
C. Fort and M. Inguscio, Phys. Rev. Lett. {\bf 93}, 140406 (2004).
\bibitem{snake} Z. Dutton, M. Budde, C. Slowe and  L.V. Hau, Science {\bf 293}, 663 (2001).
\bibitem{notef2} Clearly, for a symmetrical condensate, $\int dv_x F(v_x-\overline{v}_x) > 0$.

\end{thebibliography}
\end{document}